\documentclass[aps,pre,reprint,superscriptaddress,floatfix]{revtex4-2}
\usepackage[T1]{fontenc}
\usepackage{lmodern}
\usepackage{amsmath,amssymb,graphicx,xcolor}
\usepackage{tikz}
\usetikzlibrary{arrows.meta}
\usepackage[hidelinks]{hyperref}
\newcommand{\E}{\mathbb E}
\newcommand{\Tr}{\operatorname{Tr}}
\newcommand{\ddk}{\frac{dk}{2\pi}}
\newcommand{\ddp}{\frac{dp}{2\pi}}
\definecolor{tailblue}{HTML}{246A8B}
\definecolor{edgeorange}{HTML}{B6652E}
\definecolor{softblue}{HTML}{EAF3F7}
\definecolor{softorange}{HTML}{FBF0E6}

\begin{document}
\title{Long-Range Correlated Random Matrices}
\author{Abbas Ali Saberi}
\email{asaberi@constructor.university}
\affiliation{School of Science, Constructor University, Campus Ring 1, 28759 Bremen, Germany}
\affiliation{Max Planck Institute for the Physics of Complex Systems, 01187 Dresden, Germany}
\author{Roderich Moessner}
\affiliation{Max Planck Institute for the Physics of Complex Systems, 01187 Dresden, Germany}

\begin{abstract}
Motivated by the importance ascribed to correlations in random matrices used to model phenomena in various scientific disciplines, we report how algebraic correlations between matrix elements affect the eigenvalue statistics and spectral density of random matrices. These correlations, introduced through thresholded Gaussian fields on a square lattice, decay as a power law $\propto r^{-2H}$, with exponent $H>0$. As $H$ varies, both the eigenvalue distribution and excess kurtosis undergo qualitative changes. The spectral variance approaches $1/2$, while the moment of order $2n$ is finite only for $H>(n-1)/(2n)$. Long-wavelength scaling gives an algebraic density tail $|\lambda|^{-1-2/(1-2H)}$ for $0<H<1/2$ and the corresponding finite-size powers and logarithms. At $H=1/2$ the support is unbounded although every fixed moment is finite; for $H>1/2$ it is bounded and approaches the semicircle as $H\to\infty$. A self-consistent equation determines the limiting density from the binary covariance and fixes the excess kurtosis through the squared correlations along a row. Our results link real-space correlations to spectral regimes, distinguishing the roles of the long-distance exponent and the full covariance.
\end{abstract}
\maketitle

\textit{Introduction.}---How do long-range correlations between their matrix elements affect the statistics of eigenvalues in random matrices? This question touches upon the foundations of random matrix theory (RMT), a field that has traditionally focused on matrices with uncorrelated entries \cite{guhr1998,mehta2004random}. RMT provides a statistical description of collective behavior in systems with many fluctuating degrees of freedom. For real symmetric matrices with independent, identically distributed off-diagonal entries of finite variance, the Wigner semicircle is the basic global reference \cite{wigner1958distribution,dyson1962}. Spatial structure introduces another source of collective behavior: a correlated patch can influence many entries at once.

In many structured systems, matrix elements are correlated through geometry or an underlying field; this motivates a systematic study of random matrices with controlled entry correlations. Matrix representations of percolation connect collective eigenvalues to critical finite-size scaling \cite{saber2022universal,malekan2022exact}. Matrices drawn from statistical-mechanical configurations inherit interaction-induced correlations \cite{saberi2024interaction}, while overlap spectra in spin glasses exhibit a density crossover described by $q$-Gaussian shapes \cite{onder2026q}. These examples raise a common physical question: how is the spatial organization of bounded variables expressed in their collective spectrum?

A variety of correlated ensembles have been proposed to model structured systems. Euclidean random matrices encode geometry directly, and correlated Wishart ensembles describe covariance measurements with structured inputs \cite{mezard1999spectra,burda2005spectral}. Translation-invariant (stationary) random fields, for which correlations depend only on the separation between lattice sites and not on their absolute positions, provide a direct setting for correlations between matrix entries \cite{anderson2008,banna2015,chakrabarty2016}. Their limiting spectral measures can be related to the Fourier transform of this separation-dependent covariance, which resolves the correlations into lattice wavevectors, or momentum modes, including fields with long-range dependence, where correlations decay slowly with spatial separation rather than becoming negligible over a finite range \cite{peligrad2016,peligrad2018}. More general correlated ensembles are described by matrix Dyson equations \cite{ajanki2019,erdos2019slow}. A related lattice construction uses Boltzmann weights to connect real-space correlations to a momentum-dependent spectral response \cite{onder2026boltzmann}. The task here is to extract explicit spectral regimes from such a response when the spatial correlations decay algebraically.

Broad spectra also occur in L\'evy matrices, generalized invariant ensembles, and superstatistical mixtures \cite{cizeau1994,Toscano2004,Bertuola2004,beck2003,AbulMagd2005,BohigasCarvalhoPato2008}. In the present construction every unscaled matrix element is bounded. Large eigenvalues instead arise from collective fluctuations involving many correlated entries. Thus broad spectra here do not originate from individually heavy-tailed matrix elements, but from their spatial organization. The same long-wavelength fluctuations can have little effect on the variance and a large effect on higher moments, which place progressively more weight on the spectral tails.

Here, we address this problem by constructing random matrix ensembles from
correlated binary configurations, generated via thresholded Gaussian fields
with algebraic correlations $\sim r^{-2H}$
\cite{weinrib1984,prakash1992,makse1996method}. This introduces tunable
spatial correlations while preserving symmetry and zero ensemble mean. We
find a hierarchy of moment thresholds {where the threshold $H_n$ for each even moment $\mu_{2n}$
separates a regime in which that moment remains finite as $L\to\infty$ from
one in which it grows with system size.} These thresholds accumulate at
$H=1/2$, where the limiting support {---the range of eigenvalues over which the limiting density is
nonzero---} changes from unbounded to bounded. The long-distance correlation
exponent determines the algebraic spectral tail and the associated
finite-size growth of the divergent moments. The complete binary covariance,
including correlations at short and intermediate distances, determines the
finite moments and the central density, {i.e., the shape of the eigenvalue density around the center of
the spectrum where most eigenvalues lie.} Figure~\ref{fig:regimes} summarizes
these results; the following sections develop the physical mechanism and the
numerical tests in turn.

\begin{figure}[tb]
\centering
\begin{tikzpicture}[x=1cm,y=1cm,font=\fontsize{8}{9}\selectfont,
  line cap=round,line join=round,>=Latex]
\path[use as bounding box] (-0.10,-2.01) rectangle (8.42,2.26);
\node at (4.08,2.07) {$\mu_2=1/2\qquad(H>0)$};
\fill[softblue,rounded corners=2pt] (0,1.05) rectangle (4.80,1.81);
\fill[softorange,rounded corners=2pt] (4.80,1.05) rectangle (8.10,1.81);
\node[text=tailblue,font=\bfseries\fontsize{8}{9}\selectfont] at (2.40,1.59) {Algebraic tails};
\node[text=tailblue] at (2.40,1.27) {$\mathcal P_H(\lambda)\sim |\lambda|^{-1-2/(1-2H)}$};
\node[text=edgeorange,font=\bfseries\fontsize{8}{9}\selectfont] at (6.45,1.59) {Bounded support};
\node[text=edgeorange] at (6.45,1.27) {$\mathcal P_H\to\mathcal P_{\rm sc}\quad(H\to\infty)$};
\draw[tailblue,line width=.8pt] (4.80,.83)--(4.80,1.85);
\draw[-{Latex[length=1.5mm]},line width=.65pt] (0,.86)--(8.18,.86);
\node[right,inner sep=1pt] at (8.19,.86) {$H$};
\foreach \x in {0,2.40,4.80}{
  \draw (\x,.81)--(\x,.91);
}
\node[below,inner sep=2pt] at (0,.81) {$0$};
\node[below,inner sep=2pt] at (2.40,.81) {$1/4$};
\node[below,inner sep=2pt] at (4.80,.81) {$1/2$};
\fill[edgeorange] (6.88,.86) circle (1.45pt);
\node[below,inner sep=2pt,text=edgeorange] at (6.88,.81) {$H_0$};
\draw (7.66,.81)--(7.73,.91);
\draw (7.76,.81)--(7.83,.91);
\node[below,inner sep=2pt] at (7.99,.81) {$\infty$};
\node[text=tailblue,align=center,font=\fontsize{7}{8}\selectfont] at (4.75,.22)
 {unbounded support\\all moments finite};
\node[text=edgeorange,font=\fontsize{7}{8}\selectfont] at (7.29,.21) {$K_{H_0}=0$};
\draw[gray!25,line width=.45pt] (0,-.07)--(8.10,-.07);
\node[anchor=west,font=\bfseries\fontsize{8}{9}\selectfont] at (0,-.33) {Moment thresholds};
\node[anchor=east] at (8.10,-.33) {$H_n=(n-1)/(2n)$};
\node at (.72,-.75) {$\mu_4$};
\node at (2.39,-.75) {$\mu_6$};
\node at (4.06,-.75) {$\mu_8$};
\node at (5.53,-.93) {$\cdots$};
\node at (7.04,-.75) {$\mu_{2n}$};
\node[text=tailblue] at (.72,-1.08) {$1/4$};
\node[text=tailblue] at (2.39,-1.08) {$1/3$};
\node[text=tailblue] at (4.06,-1.08) {$3/8$};
\node[text=tailblue] at (7.04,-1.08) {$H_n\nearrow1/2$};
\draw[gray!25,line width=.45pt] (0,-1.31)--(8.10,-1.31);
\node[font=\fontsize{7}{8}\selectfont] at (1.35,-1.56) {$H<H_n$};
\node[font=\fontsize{7}{8}\selectfont] at (4.05,-1.56) {$H=H_n$};
\node[font=\fontsize{7}{8}\selectfont] at (6.75,-1.56) {$H>H_n$};
\node[font=\fontsize{7}{8}\selectfont] at (1.35,-1.88) {$m_{2n}(L)\sim L^{n(1-2H)-1}$};
\node[font=\fontsize{7}{8}\selectfont] at (4.05,-1.88) {$m_{2n}(L)\sim\log L$};
\node[font=\fontsize{7}{8}\selectfont] at (6.75,-1.88) {$\mu_{2n}<\infty$};
\end{tikzpicture}
\caption{Spectral regimes generated by algebraic entry correlations.
The variance remains finite throughout the hierarchy. The moment thresholds
accumulate at the support boundary $H=1/2$. For the covariance in
Eq.~\eqref{eq:rho}, the excess kurtosis vanishes at
$H_0=0.716556\ldots$. The moment-threshold sequence shown in the lower part of the figure is schematic.}
\label{fig:regimes}
\end{figure}

\textit{Model.}---We use the binary-field construction of long-range correlated site percolation on a square lattice. Site occupancies are interpreted as matrix elements according to their geometric positions. We start with a stationary, zero-mean, unit-variance Gaussian field $h_{ij}$ with covariance
\begin{equation}
 \rho_H(r,s)=\E[h_{ij}h_{i+r,j+s}]
 =(1+r^2+s^2)^{-H},\qquad H>0.
 \label{eq:rho}
\end{equation}
Distances are measured in lattice units. At large separation the covariance decays as $(r^2+s^2)^{-H}$. Its value at short distances is specified as well, since those correlations enter the finite spectral moments. Smaller $H$ therefore means more slowly decaying, and hence longer-ranged, correlations.

The field is converted into occupancy variables: $X_{ij}=+1$ if $h_{ij}\leq0$ and $X_{ij}=-1$ otherwise. Each sign has ensemble probability $1/2$. We ensure real eigenvalues by symmetrizing and use the usual matrix-size normalization,
\begin{equation}
 \mathcal M=\frac{X+X^T}{2},\qquad
 A_L=\frac{\mathcal M}{\sqrt L}.
 \label{eq:model}
\end{equation}
The off-diagonal elements of $\mathcal M$ are in $\{-1,0,+1\}$ and its diagonal elements are $\pm1$. The exact Gaussian sign identity \cite{vanvleck1966} gives their underlying binary covariance,
\begin{equation}
 R_H(r,s)=\E[X_{ij}X_{i+r,j+s}]
 =\frac{2}{\pi}\arcsin\rho_H(r,s).
 \label{eq:arcsine}
\end{equation}
Since $\arcsin\rho=\rho+O(\rho^3)$, the binary field has the same long-distance exponent: $R_H(r,s)\sim(2/\pi)(r^2+s^2)^{-H}$. Thresholding therefore changes the correlations at short distances but leaves their long-distance power law unchanged.

A global reversal of the field sends $A_L\to-A_L$, making the
ensemble-averaged spectral density even. We define the pooled moments as
$m_p(L)=L^{-1}\E\Tr A_L^p$, where the average is over independent
realizations of the full $L\times L$ random matrix $A_L$, with the
normalization in Eq.~\eqref{eq:model}. The limiting density is denoted by
$\mathcal P_H$, and its moments by
$\mu_p=\int\lambda^p\mathcal P_H(\lambda)\,d\lambda$ when finite.
We use $m_p(L)$ for a finite matrix and $\mu_p$ for its $L\to\infty$
limit when that limit is finite.

\textit{Scaling analysis of eigenvalue statistics.}---We analyze the eigenvalue moments to characterize the spectrum. The variance follows directly from the squared matrix entries. Pairing each entry with its transpose gives the exact finite-size identity
\begin{equation}
 m_2(L)=\frac12+\frac{1}{2L}
 +\frac{1}{L^2}\sum_{r=1}^{L-1}(L-r)R_{H,L}(r,-r),
 \label{eq:m2exact}
\end{equation}
where $R_{H,L}$ is the binary covariance produced by the finite periodic $L\times L$ field construction used in the simulations.
The last term measures correlations between transpose partners.
These are the entries $X_{ij}$ and $X_{ji}$ combined by the symmetrization in Eq.~\eqref{eq:model}.
For an $L\times L$ restriction of the infinite stationary field,
without periodic identification, this term decays as $L^{-2H}$
for $0<H<1/2$, as $L^{-1}\log L$ at $H=1/2$, and as $L^{-1}$
for $H>1/2$. Thus $m_2(L)\to1/2$. The exact covariance and finite-size conventions
for the periodic construction used numerically are given in the
Supplemental Material~\cite{SM}, Sec.~S1.

Higher moments resolve how the real-space covariance is distributed over momentum modes. We write the binary covariance as
\begin{equation}
 R_H(r,s)=\int_{-\pi}^{\pi}\!\ddk\int_{-\pi}^{\pi}\!\ddp\,
 e^{i(kr+ps)}F_H(k,p).
 \label{eq:FH}
\end{equation}
The nonnegative function $F_H$ is the Fourier-space spectral density of the field covariance.
Here $k$ and $p$ are the wavevectors conjugate to the two lattice separations $r$ and $s$; below we refer to them as momentum modes. The value $F_H(k,p)$ specifies the covariance weight associated with that pair of modes.
Small momenta correspond to long wavelengths, so the behavior of $F_H(k,p)$ near $(0,0)$ encodes the long-distance correlations.
For this stationary field, the limiting matrix spectrum is described by \cite{peligrad2016,peligrad2018}
\begin{equation}
 g_H(k,z)=-\left[z+\frac12\int_{-\pi}^{\pi}\!\ddp\,
 F_H(k,p)g_H(p,z)\right]^{-1}.
 \label{eq:dyson}
\end{equation}
Here $g_H(k,z)$ is the resolvent resolved by momentum $k$.
For $\operatorname{Im}z>0$, the physical solution has
$\operatorname{Im}g_H>0$ and $g_H\sim-1/z$ at large $z$.
The integral gives the covariance-weighted feedback on mode $k$
from all momentum modes $p$: each response $g_H(p,z)$ enters with weight
$F_H(k,p)/2$.
Equation~\eqref{eq:dyson} is therefore the bridge between the real-space correlations and the eigenvalue spectrum.
The field is a bounded function of filtered independent Gaussian variables,
so the stationary-field spectral result applies. Symmetrization gives the
kernel $[F_H(k,p)+F_H(p,k)]/4=F_H(k,p)/2$ in our normalization; the steps
are detailed in Supplemental Material, Sec.~S2.

Averaging over momentum gives the usual resolvent and hence the eigenvalue density,
\begin{equation}
 \begin{aligned}
 G_H(z)&=\int_{-\pi}^{\pi}\!\ddk\,g_H(k,z),\\
 \mathcal P_H(\lambda)&=\frac1\pi\lim_{\eta\downarrow0}
 \operatorname{Im}G_H(\lambda+i\eta).
 \end{aligned}
 \label{eq:densityfromg}
\end{equation}
Once the binary covariance is specified, these equations determine the limiting eigenvalue density of $A_L$.

Expanding the same equation at large $|z|$ generates the spectral moments; the derivation is given in Supplemental Material, Sec.~S2. For the present argument, the first two even moments are sufficient:
\begin{equation}
 \mu_2=\frac12,\qquad
 \mu_4=\frac12\sum_{r\in\mathbb Z}R_H(r,0)^2.
 \label{eq:m4}
\end{equation}
The fourth moment measures the total squared correlation along a row. Since $R_H(r,0)^2\sim r^{-4H}$, it is finite for $H>1/4$ and diverges logarithmically at $H=1/4$. Thus the variance can remain finite while a higher moment diverges: higher moments are more sensitive to the rare large eigenvalues generated by long-range correlations.

The same idea extends to all even moments. For $0<H<1$, the long-wavelength covariance behaves as
$F_H(k,p)\propto(k^2+p^2)^{H-1}$.
The most infrared-sensitive contribution to $\mu_{2n}$ contains $n$ covariance factors and $n+1$ momentum integrations. If all momenta lie at a common small scale $\kappa$, a logarithmic momentum shell contributes
\begin{equation*}
 \kappa^{\,1-n(1-2H)}\,\frac{d\kappa}{\kappa}.
\end{equation*}
The small-$\kappa$ integral converges only when $1-n(1-2H)>0$. This gives the thresholds
\begin{equation}
 \mu_{2n}<\infty\quad\Longleftrightarrow\quad
 H>H_n\equiv\frac{n-1}{2n},\qquad n\geq2.
 \label{eq:thresholds}
\end{equation}
The term-by-term derivation of this infrared counting is given in Supplemental Material, Sec.~S2.
At $H=H_n$, the shell contribution is proportional to $d\kappa/\kappa$, so equal logarithmic intervals of wavelength contribute equally, giving the finite-size growth $m_{2n}(L)\sim a_n\log L$.

\begin{figure}[tb]
\centering
\includegraphics[width=\columnwidth]{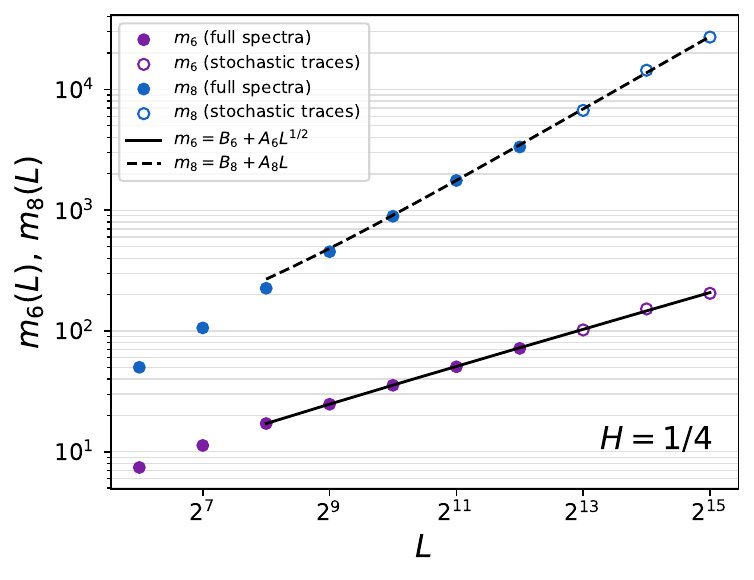}
\caption{Moment hierarchy at $H=1/4$. Symbols show $m_6(L)$ (purple)
and $m_8(L)$ (blue) for $L=2^6,\ldots,2^{15}$. Filled circles denote
full-spectrum measurements for $L\leq 2^{12}$, while open circles denote
stochastic trace-based estimates for $L\geq 2^{13}$. The black solid and
dashed curves are fits for $L\geq256$ to
$B_6+A_6L^{1/2}$ and $B_8+A_8L$, respectively, with the powers fixed by
Eq.~\eqref{eq:finitepred}. The fitted coefficients are
$A_6=1.1563$, $B_6=-1.3843$, $A_8=0.8342$, and $B_8=54.85$.
Moment estimators and numerical details are given in Supplemental Material, 
Sec.~S2.}
\label{fig:hierarchy}
\end{figure}

A finite system has a largest wavelength of order $L$, or equivalently a smallest nonzero momentum of order $L^{-1}$. The corresponding finite-size scaling is
\begin{equation}
 m_{2n}(L)\sim
 \begin{cases}
 a_{n,H}L^{n(1-2H)-1},&0<H<H_n,\\
 a_n\log L,&H=H_n,
 \end{cases}
 \label{eq:finitepred}
\end{equation}
with finite backgrounds and amplitudes that depend on the finite periodic construction. At $H=1/4$, this gives $m_4\sim\log L$, $m_6\sim L^{1/2}$, and $m_8\sim L$. Figure~\ref{fig:hierarchy} compares the last two predictions with moment estimates through $L=2^{15}$. The thresholds $1/4,1/3,3/8,\ldots$ accumulate at $1/2$: progressively higher moments resolve the increasingly rare large eigenvalues. Higher moments therefore act as increasingly sensitive probes of the spectral tail.

\begin{figure}[t!]
\centering
\includegraphics[width=\columnwidth]{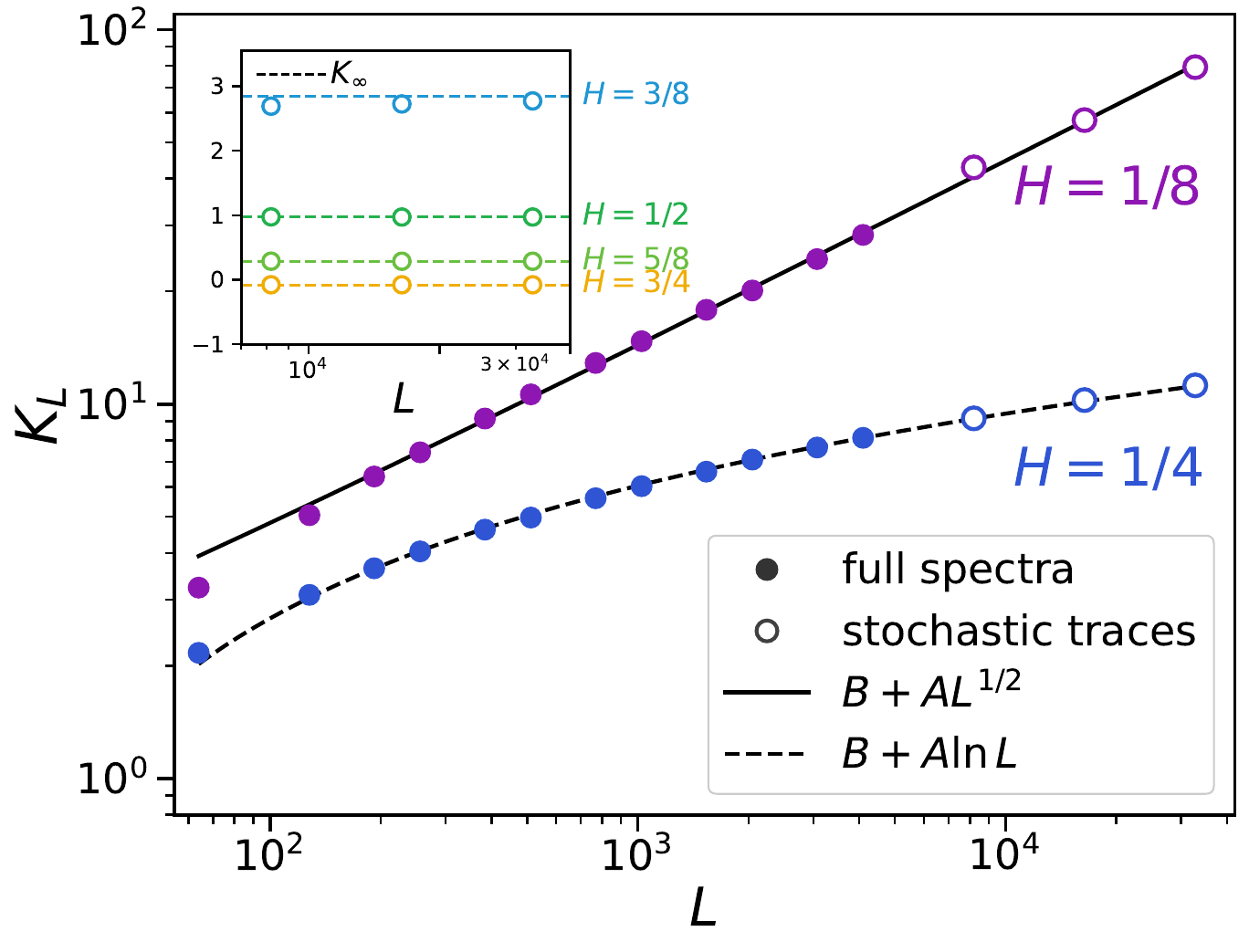}
\caption{Excess kurtosis across the correlation hierarchy.
The main panel shows $K_L$ versus $L$ for the two growing branches,
$H=1/8$ (purple) and $H=1/4$ (blue), over the full available size
range. Filled circles denote the smaller-$L$ full-spectrum
measurements, while open circles denote the large-$L$ stochastic
trace-based estimates. The black solid and dashed curves are fits to
$K_L=B+AL^{1/2}$ for $H=1/8$ and $K_L=B+A\ln L$ for $H=1/4$,
respectively, with the scaling forms fixed by
Eq.~\eqref{eq:finitepred}.
The inset focuses on the large-$L$ finite-kurtosis regime
$H=3/8,\,1/2,\,5/8$, and $3/4$; open circles show the stochastic
trace-based estimates. Colored dashed lines mark the parameter-free
asymptotic predictions of Eq.~\eqref{eq:kurtosis},
$K_\infty=2.83725$, $0.96863$, $0.28435$, and $-0.07984$
for $H=3/8$, $1/2$, $5/8$, and $3/4$, respectively.
Error bars on the stochastic trace-based estimates denote 95\%
confidence intervals over independent realizations, including
trace-probe uncertainty.}
\label{fig:EK}
\end{figure}

\textit{Spectral crossover.}---We next quantify how the eigenvalue distribution changes as the correlation exponent $H$ is varied, first through the excess kurtosis and then through higher moments. The excess kurtosis follows the changing weight of the spectral tails. We use $K_L=m_4(L)/m_2(L)^2-3$. A Gaussian has $K=0$, while the variance-$1/2$ semicircle has $K=-1$; positive $K$ therefore indicates enhanced tail weight relative to a Gaussian. For $H>1/4$, Eq.~\eqref{eq:m4} gives the limiting value
\begin{equation}
 K_H=\frac{\mu_4}{\mu_2^2}-3
 =-1+4\sum_{r=1}^{\infty}R_H(r,0)^2.
 \label{eq:kurtosis}
\end{equation}
For the covariance in Eq.~\eqref{eq:rho}, the values at $H=3/8$, $1/2$, $5/8$, and $3/4$ are $2.83725$, $0.96863$, $0.28435$, and $-0.07984$, respectively. The sum decreases monotonically with $H$, crossing zero at $H_0=0.716556\ldots$ and tending to the semicircle value $-1$. Its value and zero depend on the correlations at all separations. Figure~\ref{fig:EK} compares these parameter-free limiting values with the finite-size data.

Higher moments resolve the spectral shape further. At $H_0$, where the fourth cumulant vanishes, direct evaluation of the next nontrivial moment gives $\mu_6\simeq1.7271$. A Gaussian with the same variance has $\mu_6=15/8=1.875$. Equivalently, the sixth cumulant at the kurtosis zero is $\kappa_6=\mu_6-15/8\simeq-0.1479$. Thus $K_{H_0}=0$ does not make the spectrum Gaussian: the sixth cumulant still distinguishes the two distributions. The corresponding covariance sums are given in Supplemental Material, Sec.~S2.

The support boundary, meaning $H=1/2$, where the limiting eigenvalue density changes from unbounded support for $H\leq1/2$ to bounded support for $H>1/2$, has a separate physical origin. For $H>1/2$, the row correlation is summable, i.e., $\sum_r R_H(r,0)$ is finite. The total covariance weight seen by every momentum mode is then bounded, and Eq.~\eqref{eq:dyson} gives a bounded limiting spectrum. At $H=1/2$, the small-momentum covariance weight instead grows logarithmically. Every fixed moment remains finite, but sufficiently high moments grow at least factorially with their order, which excludes bounded support. Thus $H=1/2$ is unusual: all moments of any fixed order exist, although the limiting density still extends to arbitrarily large $|\lambda|$. As $H\to\infty$, off-site correlations vanish and Eq.~\eqref{eq:dyson} reduces to the semicircle on $[-\sqrt2,\sqrt2]$. The corresponding bounds are given in Supplemental Material, Sec.~S2.

\textit{Eigenvalue density and tail.}---Equation~\eqref{eq:dyson} determines the one-point density of the eigenvalues of $A_L$ in the $L\to\infty$ limit throughout the spectrum. Its tail for $0<H<1/2$ follows from the same long-wavelength physics as the moment hierarchy. Write $\alpha=1-2H>0$. At small momentum,
$F_H(k,p)\sim(k^2+p^2)^{-(1+\alpha)/2}$. For modes with
$|k|,|p|\sim\kappa$, the covariance term in Eq.~\eqref{eq:dyson}
therefore scales as $\kappa^{-\alpha}g$. In the far spectral tail
$g\sim1/\lambda$, so this term becomes comparable to $\lambda$ when
\begin{equation*}
\lambda^2\sim\kappa^{-\alpha},
\qquad
\kappa\sim|\lambda|^{-2/\alpha}.
\end{equation*}
Only a fraction of order $\kappa$ of the momentum modes lies in this long-wavelength region, while each contributes a resolvent of order $1/|\lambda|$. Hence
\begin{equation}
 \mathcal P_H(\lambda)\sim
 C_H|\lambda|^{-1-f(H)},\qquad
 f(H)=\frac{2}{1-2H},
 \label{eq:tail}
\end{equation}
for $0<H<\frac12$. The tail amplitude $C_H$ depends on the full covariance, while the exponent depends only on its long-distance power law. The rescaled form of Eq.~\eqref{eq:dyson} behind this asymptotic balance is given in Supplemental Material, Sec.~S2.

The smallest nonzero momentum in a finite sample is of order $L^{-1}$, giving the characteristic large-eigenvalue scale $\lambda_L\sim L^{1/2-H}$. This is where the infinite-system algebraic tail is cut off by finite size. Integrating $\lambda^{2n}\mathcal P_H(\lambda)$ up to this scale yields $L^{n(1-2H)-1}$ when $2n>f(H)$ and $\log L$ when $2n=f(H)$. Thus the same long-wavelength cutoff that produces the algebraic tail also produces the power-law or logarithmic finite-size growth of the divergent moments. The scale $\lambda_L$ describes the finite-size tail region; sample-to-sample fluctuations determine individual extreme eigenvalues.

\begin{figure}[tb]
\centering
\includegraphics[width=\columnwidth]{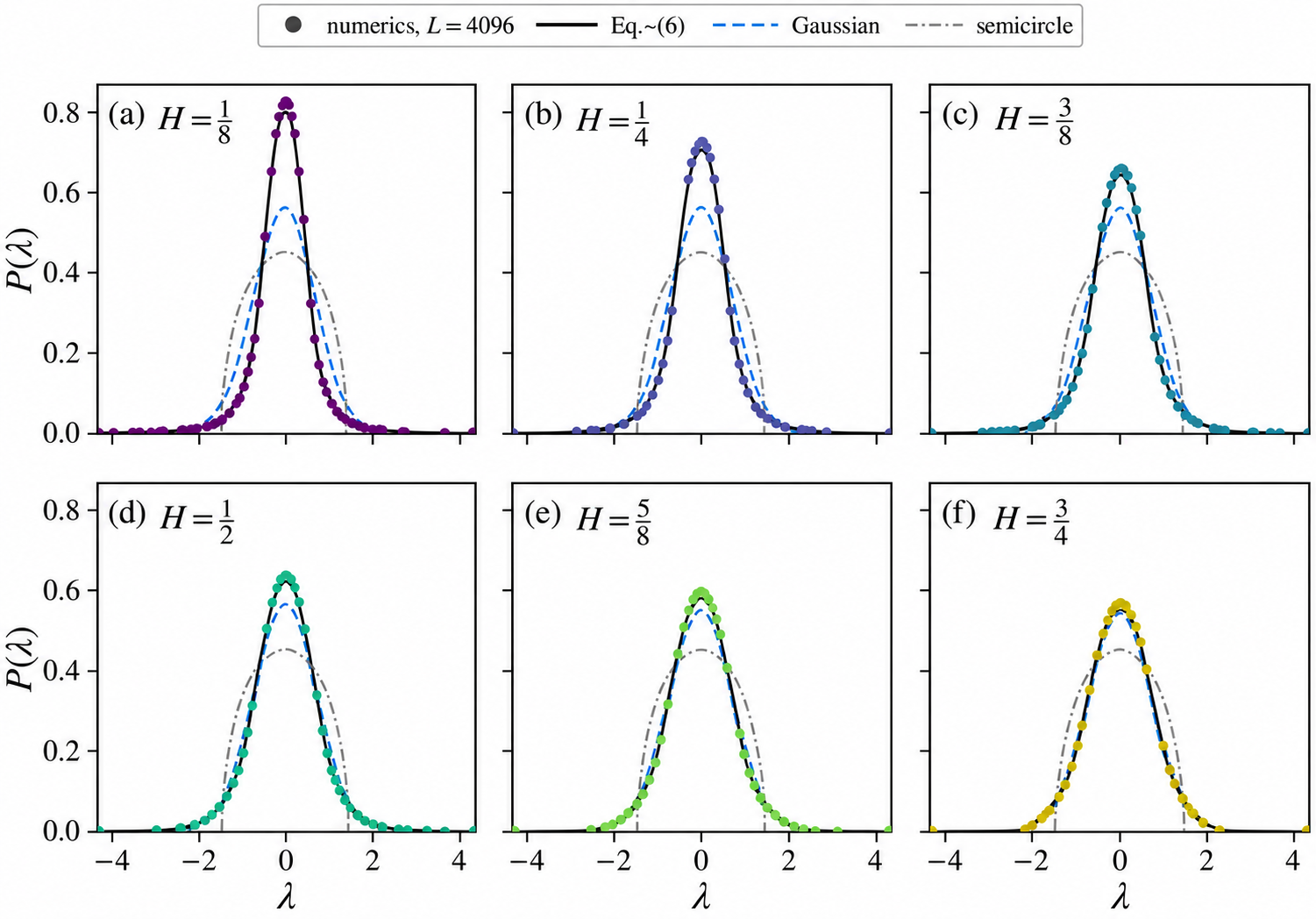}
\caption{One-point spectral densities at $L=4096$ for
$H=1/8,1/4,3/8,1/2,5/8$, and $3/4$. Colored symbols show the numerical
densities; solid black curves show the covariance equation using the
binary covariance of the finite periodic construction and Lorentzian resolution
$\eta=0.02$. Gaussian and semicircle curves, both with variance $1/2$,
provide fixed reference shapes. The panels resolve the central
density; the tail scaling is given by Eq.~\eqref{eq:tail}.}
\label{fig:density}
\end{figure}

The tail can also be expressed in the notation of $q$-Gaussian distributions, $\mathcal P_q(\lambda)\propto[1+(q-1)b\lambda^2]^{-1/(q-1)}$ \cite{tsallis1988,Toscano2004,Bertuola2004}. Matching the asymptotic decay exponents gives
\begin{equation}
 q_{\rm tail}(H)=1+\frac{2}{f(H)+1}
 =\frac{5-6H}{3-2H},\qquad 0<H<1/2.
 \label{eq:qtail}
\end{equation}
This parameter encodes only the asymptotic tail exponent. It therefore serves as a convenient coordinate for the tail rather than for the full eigenvalue density. The central shape is obtained from the complete covariance in Eq.~\eqref{eq:dyson}. For example, at $H=3/8$, the tail has $f=8$, corresponding to a $q$-Gaussian excess kurtosis $3/2$, while the matrix covariance gives $K_H=2.83725$. Thus the long-distance exponent fixes the asymptotic tail, whereas correlations at all separations determine the central density and its kurtosis.
Figure~\ref{fig:density} compares the finite-$L$ densities directly with the solution of the covariance equation.

\textit{Numerical simulations.}---Independent Gaussian fields are generated by Fourier filtering \cite{makse1996method} on a periodic $L\times L$ lattice. Independent Gaussian noise is filtered in momentum space so that the resulting real-space field has the desired covariance. The filtering spectrum, zero-mode convention, and normalization are specified in Supplemental Material, Sec.~S1. The binary covariance used in the finite-size calculations is obtained from the actual filtering spectrum through Eq.~\eqref{eq:arcsine}. Figure~S1 in Supplemental Material illustrates the Gaussian and binary correlation decay for the prescribed infinite-lattice covariance.

Full eigenspectra were obtained for $L=2^6,\ldots,2^{12}$,
with the number of independent matrices decreasing from $10^4$
at the smallest size to $500$ at $L=4096$. At the larger sizes
$L=8192,16384,32768$, we used respectively $1000$, $500$, and
$300$ independent matrix realizations for each value of $H$.
The second moment was evaluated exactly, while the fourth and,
where required, higher even moments were estimated with independent
Rademacher trace probes. These probes estimate traces without diagonalizing the matrix and therefore make the larger sizes accessible. The fourth-moment estimates used 32 probes per matrix for $H=1/8,\,1/2,\,5/8$, and $3/4$, and 64 probes per matrix for $H=1/4$ and $3/8$. At $H=1/4$, the sixth and eighth moments were obtained from the same 64-probe calculations. The estimators, calibration against diagonalization, and fit coefficients are collected in Supplemental Material, Sec.~S2. Pooled moments give equal weight to each matrix realization;
uncertainty estimates resample the independently generated matrices
as whole statistical samples and include trace-probe fluctuations.

Figures~\ref{fig:hierarchy} and \ref{fig:EK} examine the predicted power-law and logarithmic growth of the moments, while Fig.~\ref{fig:density} tests the covariance equation for the central density. Because Lorentzian broadening generates an artificial $|\lambda|^{-2}$ far tail, the unsmoothed cumulative eigenvalue count defined in Supplemental Material, Sec.~S2, provides a direct measurement of the empirical tail without this broadening artifact.

\textit{Conclusions.}---Long-range correlations between bounded matrix entries produce a hierarchy of collective spectral fluctuations. The variance remains $1/2$ in the large-matrix limit, while higher moments become sensitive to increasingly long wavelengths. Their thresholds $H_n=(n-1)/(2n)$ accumulate at $H=1/2$, separating unbounded from bounded support of the limiting eigenvalue density as $L\to\infty$. Long-wavelength scaling gives the algebraic tail below this boundary and the associated finite-size powers and logarithms. The complete binary covariance fixes the limiting density, its finite kurtosis, and higher shape measures such as the sixth cumulant. The central distinction is therefore between what is fixed by the asymptotic decay of correlations---the moment thresholds and tail exponent---and what requires the covariance at all distances---the detailed spectral shape and its finite moments. These results connect techniques from statistical physics, correlated percolation, and random matrix analysis through a direct relation between real-space correlation structure and spectral fluctuations.

\textit{Acknowledgments.}---This work was funded by the Deutsche Forschungsgemeinschaft (DFG, German Research Foundation) under Project No.~557852701 (A.A.S.), and in part by the Deutsche Forschungsgemeinschaft via Research Unit FOR 5522 (project-id 499180199) and the cluster of excellence ctd.qmat (EXC 2147, Project No.~390858490) (R.M.). The study was also supported by the Advanced Study Group
``Strongly Correlated Extreme Fluctuations'' at the Max Planck Institute for
the Physics of Complex Systems, Dresden (2024/25) \cite{pks_asg2024}.

\section*{Data Availability}

The data that support the findings of this article are not publicly available.
The data are available from the authors upon reasonable request.

\bigskip

\onecolumngrid

\begin{center}
{\large\bf Supplemental Material for\\
``Long-Range Correlated Random Matrices''}

\vspace{0.8em}

Abbas Ali Saberi$^{1,2,*}$ and Roderich Moessner$^{2}$

\vspace{0.4em}

{\small
$^{1}$School of Science, Constructor University,
Campus Ring 1, 28759 Bremen, Germany\\
$^{2}$Max Planck Institute for the Physics of Complex Systems,
01187 Dresden, Germany\\
$^{*}$asaberi@constructor.university
}
\end{center}

\vspace{1em}

\setcounter{section}{0}
\setcounter{equation}{0}
\setcounter{figure}{0}

\renewcommand{\thesection}{S\arabic{section}}
\renewcommand{\theequation}{S\arabic{equation}}
\renewcommand{\thefigure}{S\arabic{figure}}

\renewcommand{\theHsection}{supp.S\arabic{section}}
\renewcommand{\theHequation}{supp.S\arabic{equation}}
\renewcommand{\theHfigure}{supp.S\arabic{figure}}

\twocolumngrid

This Supplemental Material provides the details behind the
physical arguments in the main text. We follow the same sequence as in the main
text: construction of the correlated field, its connection to the eigenvalue
density, the hierarchy of spectral moments, the support boundary, the
large-eigenvalue tail, and finally the numerical estimators. Matrix entries and
eigenvalues use the normalization of Eq.~(2) of the main text throughout.

\section{Field construction and exact spectral variance}
\label{sec:sm-generator}

\textit{Stationary covariance.}---The covariance in Eq.~(1) of the main text can be
written as
\begin{equation}
 (1+r^2+s^2)^{-H}
 =
 \frac{1}{\Gamma(H)}
 \int_0^\infty
 u^{H-1}e^{-u}e^{-u(r^2+s^2)}\,du .
 \label{eq:sm-mixture}
\end{equation}
This representation has a simple physical meaning: the
power-law covariance is a positive superposition of Gaussian correlation
profiles of different spatial ranges. Since each Gaussian profile has a
nonnegative Fourier spectrum, the full covariance can be generated by
filtering independent Gaussian noise.

\textit{Thresholding.}---For two unit Gaussian variables $U$ and $V$ with
correlation coefficient $\rho$, the correlation of their signs is 
\cite{vanvleck1966}
\begin{equation}
 \E[\operatorname{sgn}U\,\operatorname{sgn}V]
 =
 \frac{2}{\pi}\arcsin\rho .
 \label{eq:sm-sign}
\end{equation}
Therefore, at large separation,
\begin{equation}
 R_H(r,s)
 =
 \frac{2}{\pi}\rho_H(r,s)+O(\rho_H^3).
\end{equation}
Thresholding changes the covariance at short distances but
does not change its leading long-distance decay $r^{-2H}$. This distinction
will be important below: the long-distance power controls the spectral
exponents, while the covariance at all distances controls quantities such as
the finite kurtosis and the detailed central density.

\textit{Finite periodic construction.}---
The numerical simulations use a periodic $L\times L$ lattice.
For $r=0,\ldots,L-1$, define the periodic distance
\begin{equation*}
 d_L(r)=\min(r,L-r)
\end{equation*}
and the corresponding finite-size target covariance
\begin{equation}
 C^{(0)}_{H,L}(r,s)
 =
 \left[1+d_L(r)^2+d_L(s)^2\right]^{-H}.
 \label{eq:sm-periodic-target}
\end{equation}
The minimum-image distance is continuous but has a cusp at the antipodal point $r=L/2$ for even $L$. We therefore use Eq.~\eqref{eq:sm-periodic-target} as the finite-size target covariance, compute its discrete Fourier spectrum explicitly, set any negative Fourier components to zero in Eq.~\eqref{eq:sm-Q}, and use the resulting covariance in Eq.~\eqref{eq:sm-finitecov} in all finite-$L$ comparisons. The long-wavelength exponents are obtained from the infinite-lattice covariance, Eq.~(1) of the main text.
Its discrete Fourier transform is denoted by
$\widetilde C^{(0)}_{H,L}(\mathbf k)$, with
$\mathbf k=(2\pi a/L,2\pi b/L)$. We use the filtering spectrum
\begin{equation}
 Q_{H,L}(\mathbf k)=
 \begin{cases}
 0,&\mathbf k=0,\\
 \max\!\left[
 \operatorname{Re}\widetilde C^{(0)}_{H,L}(\mathbf k),0
 \right],&\mathbf k\neq0 .
 \end{cases}
 \label{eq:sm-Q}
\end{equation}
Independent Gaussian noise is Fourier transformed, multiplied
mode by mode by $\sqrt{Q_{H,L}}$, and transformed back to real space. This is
the Fourier-filtering construction used in the simulations. Removing the zero-momentum
mode fixes the spatial mean, while keeping $Q_{H,L}\geq0$ guarantees a valid
finite-size covariance.

The resulting normalized Gaussian covariance is
\begin{equation}
 \rho_{H,L}(r,s)
 =
 \frac{
 \sum_{\mathbf k}
 Q_{H,L}(\mathbf k)
 e^{i\mathbf k\cdot(r,s)}
 }{
 \sum_{\mathbf k}Q_{H,L}(\mathbf k)
 },
 \label{eq:sm-finite-rho}
\end{equation}
and after thresholding the binary covariance is
\begin{equation}
 R_{H,L}(r,s)
 =
 \frac{2}{\pi}\arcsin\rho_{H,L}(r,s).
 \label{eq:sm-finitecov}
\end{equation}
These finite-$L$ covariances are used whenever we compare
analytical predictions directly with the simulations. They include the finite-size changes produced by periodicity and by removal of the zero mode, while reproducing the intended algebraic decay over separations below the periodic wrap-around scale.

\begin{figure*}[tb]
\centering
\includegraphics[width=0.78\textwidth]{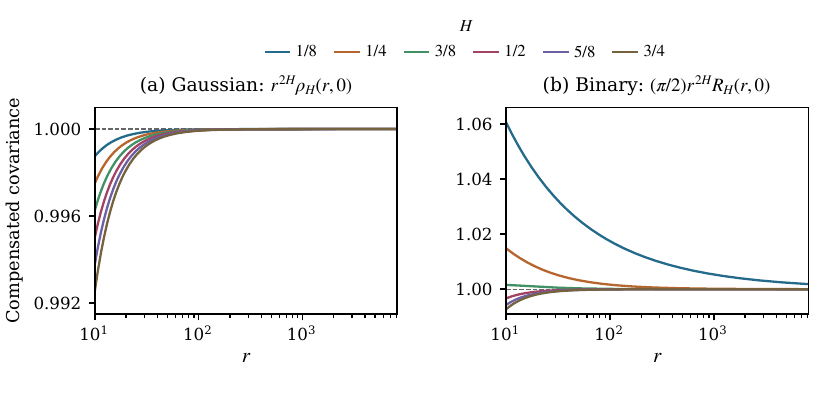}
\caption{Compensated covariances of the prescribed infinite-lattice
field. Here ``compensated'' means that the covariance is multiplied by the inverse of its expected asymptotic $r^{-2H}$ decay, including the asymptotic prefactor, so that the predicted power law appears as a plateau at unity. Both panels are direct evaluations of Eqs.~(1) and (3) of the main text. The Gaussian and binary covariances approach
the same unit plateau after compensation for their common power law.
The stronger finite-distance correction of the binary covariance at
small $H$ follows from the higher powers in the arcsine relation.}
\label{fig:sm-generator}
\end{figure*}

\textit{Second moment.}---Because $X_{ij}^2=1$, the spectral variance can be
evaluated directly:
\begin{equation}
 \begin{aligned}
 m_2(L)
 &=
 \frac{1}{4L^2}
 \sum_{i,j=1}^L
 \E\left(X_{ij}+X_{ji}\right)^2
 \\
 &=
 \frac12+\frac{1}{2L}
 +\frac{1}{L^2}
 \sum_{r=1}^{L-1}(L-r)R_{H,L}(r,-r).
 \end{aligned}
 \label{eq:sm-m2}
\end{equation}
The first term gives the limiting variance $1/2$. The
$1/(2L)$ correction comes from the diagonal, while the last term measures the
correlation between $X_{ij}$ and its transpose partner $X_{ji}$.

For an $L\times L$ square cut from the infinite field,
\begin{equation*}
 R_H(r,-r)\sim d_H r^{-2H},
 \qquad
 d_H=\frac{2}{\pi}2^{-H}.
\end{equation*}
Replacing the large-distance part of the sum by an integral gives
\begin{equation}
 m_2(L)-\frac12
 \sim
 \begin{cases}
 \displaystyle
 \frac{d_H}{(1-2H)(2-2H)}\,L^{-2H},
 &0<H<1/2,\\
 \displaystyle
 \frac{d_{1/2}\log L}{L},
 &H=1/2,\\
 \displaystyle
 \frac{\frac12+\sum_{r\ge1}R_H(r,-r)}{L},
 &H>1/2 .
 \end{cases}
 \label{eq:sm-width}
\end{equation}
All three regimes approach the same limiting value
$m_2=1/2$. Equation~\eqref{eq:sm-width} makes the asymptotic scaling
transparent, while Eq.~\eqref{eq:sm-m2} with the actual finite covariance
in Eq.~\eqref{eq:sm-finitecov} is used for the periodic simulations.

\section{Covariance response, moment hierarchy, and numerical tests}
\label{sec:sm-response}

\textit{From covariance to the density.}---
The stationary-field results of
Refs.~\cite{peligrad2016,peligrad2018} apply to stationary finite-variance
fields constructed from independent random variables. The filtered Gaussian
field and the binary field obtained from it satisfy these conditions.

For the symmetrized matrix
$A_L=(X+X^T)/(2\sqrt L)$, the covariance entering the spectral equation is
\begin{equation*}
 \frac{F_H(k,p)+F_H(p,k)}{4}.
\end{equation*}
Because our covariance is symmetric under interchange of the two lattice
directions, this reduces to $F_H(k,p)/2$, giving Eq.~(6) of the main text.

Physically, Eq.~(6) of the main text states that the spectral
response of a momentum mode $k$ is modified by all other modes $p$, with a
strength fixed by their covariance $F_H(k,p)$. Averaging over $k$ then gives
the ordinary eigenvalue density.

For a discrete momentum grid with $N$ points, the equation used numerically is
\begin{equation}
 \begin{aligned}
 g_a
 &=
 -\left[
 z+\frac{1}{2N}\sum_{b=1}^{N}F_{ab}g_b
 \right]^{-1},\\
 \mathcal P_{H,\eta}(\lambda)
 &=
 \frac{1}{\pi N}
 \sum_a\operatorname{Im}g_a.
 \end{aligned}
 \label{eq:sm-discrete-dyson}
\end{equation}
with $z=\lambda+i\eta$.
Here $F_{ab}$ is simply the discrete Fourier-space covariance.
The curves in Fig.~4 of the main text use $\eta=0.02$. Varying the momentum
grid and $\eta$ independently checks numerical discretization and spectral
broadening.

\textit{Low moments.}---
The same self-consistent equation also generates the spectral
moments by expansion at large $|z|$. A particularly useful quantity is the
total covariance weight coupled to a fixed momentum $k$, meaning the covariance weight associated with that mode after summing over all partner momenta $p$,
\begin{equation}
 v_H(k)
 =
 \frac12\int_{-\pi}^{\pi}\!\ddp\,F_H(k,p)
 =
 \frac12\sum_{r\in\mathbb Z}R_H(r,0)e^{-ikr}.
 \label{eq:sm-v}
\end{equation}
In real space this is simply the Fourier transform of the
correlations along one matrix row.

The first three even moments are
\begin{equation}
 \mu_2=\frac12,
 \qquad
 \mu_4=\frac12\sum_{r\in\mathbb Z}R_H(r,0)^2,
 \label{eq:sm-lowmoments}
\end{equation}
and
\begin{equation}
 \mu_6
 =
 2\langle v_H^3\rangle_k
 +
 \frac32
 \int_{-\pi}^{\pi}\!\ddk
 \int_{-\pi}^{\pi}\!\ddp\,
 v_H(k)F_H(k,p)v_H(p),
 \label{eq:sm-mu6}
\end{equation}
where
\begin{equation*}
 \langle f\rangle_k
 \equiv
 \int_{-\pi}^{\pi}\frac{dk}{2\pi}\,f(k).
\end{equation*}
Equation~\eqref{eq:sm-lowmoments} is the central simple
result: the fourth spectral moment is fixed by the total squared correlation
along a row. The sixth moment probes the covariance one level further,
including how different momentum modes are correlated with one another.

For the independent-entry limit, $F_H\to1$ and $v_H\to1/2$, giving
\begin{equation*}
 \mu_2=\frac12,\qquad
 \mu_4=\frac12,\qquad
 \mu_6=\frac58,
\end{equation*}
as required for the semicircle with variance $1/2$.

For the calculation at the point $H_0$ where the excess kurtosis vanishes, it is convenient to write the
sixth moment directly in real space. With
\begin{equation*}
 c_r\equiv R_H(r,0),
\end{equation*}
one finds
\begin{equation}
 \mu_6
 =
 \frac14\sum_{r,s\in\mathbb Z}c_rc_sc_{r+s}
 +
 \frac38\sum_{r,s\in\mathbb Z}
 c_rc_sR_H(r,s).
 \label{eq:sm-m6real}
\end{equation}
These sums converge for $H>1/3$. At
\begin{equation*}
 H_0=0.7165564888\ldots,
\end{equation*}
where the excess kurtosis vanishes, they give
\begin{equation*}
 \mu_6\simeq1.7271.
\end{equation*}
A Gaussian with the same variance has $\mu_6=15/8$, so
\begin{equation*}
 \kappa_6=\mu_6-\frac{15}{8}\simeq-0.1479.
\end{equation*}
Thus matching the Gaussian fourth moment at $H_0$ does not
make the full spectrum Gaussian: the sixth cumulant still detects a clear
difference. Evaluating the sums with $|r|,|s|\le4096$ gives $\mu_6=1.7270915\ldots$; increasing the cutoff from $2048$ to $4096$ changes $\mu_6$ by only $1.3\times10^{-6}$, well below the precision of the quoted value $\mu_6\simeq1.7271$.

\textit{Why the hierarchy appears.}---
For $0<H<1$, the long-distance binary covariance behaves as
$R_H(\mathbf r)\sim(2/\pi)|\mathbf r|^{-2H}$. Its small-momentum form is
\begin{equation}
 F_H(k,p)
 \sim
 A_H(k^2+p^2)^{H-1},
 \qquad
 A_H=
 2^{3-2H}\frac{\Gamma(1-H)}{\Gamma(H)}.
 \label{eq:sm-smallk}
\end{equation}
For the moment thresholds, only the exponent $H-1$ in the small-momentum singularity $(k^2+p^2)^{H-1}$ is relevant. Corrections produced by thresholding decay faster in real space
and are therefore less singular near $k=p=0$.

Consider the contribution to the moment $\mu_{2n}$ in which
all $n$ covariance factors probe the same small momentum scale $\kappa$.
There are $n+1$ momentum integrations. The available phase space therefore
contributes $\kappa^{n+1}$, while the $n$ covariance factors contribute
$\kappa^{n(2H-2)}$. A logarithmic shell of momenta consequently contributes
\begin{equation}
 \kappa^{\,1-n(1-2H)}
 \frac{d\kappa}{\kappa}.
 \label{eq:sm-degree}
\end{equation}

This one expression contains the whole hierarchy. For $H<1/2$, contributions in which only $q<n$ covariance factors are simultaneously at small momentum scale as $\kappa^{1-q(1-2H)}$ and are therefore less singular than the $q=n$ contribution above. If
$1-n(1-2H)>0$, the small-momentum integral converges. At equality it gives a
logarithm. Below equality it is dominated by the smallest available momentum.
Therefore
\begin{equation}
 H_n=\frac{n-1}{2n},
 \qquad
 \mu_{2n}<\infty
 \quad\Longleftrightarrow\quad
 H>H_n .
 \label{eq:sm-threshold}
\end{equation}
All contributions to the moment are nonnegative, so this
infrared divergence cannot be removed by cancellation with another term.

For a finite system the smallest momentum is of order $L^{-1}$. Substituting
this lower cutoff into Eq.~\eqref{eq:sm-degree} gives
\begin{equation}
 m_{2n}(L)
 \sim
 \begin{cases}
 L^{n(1-2H)-1},&H<H_n,\\
 \log L,&H=H_n,
 \end{cases}
 \label{eq:sm-finitesize}
\end{equation}
apart from finite backgrounds and nonuniversal amplitudes.
This is the finite-size scaling quoted in
Eq.~(10) of the main text.

\textit{Bounded and unbounded support.}---
For $H>1/2$, the row covariance is summable:
\begin{equation*}
 \sum_rR_H(r,0)<\infty.
\end{equation*}
Consequently,
\begin{equation*}
 v_H(k)
 =
 \frac12\sum_rR_H(r,0)e^{-ikr}
\end{equation*}
remains bounded for every $k$. Defining
\begin{equation*}
 V_H=\frac12\sum_rR_H(r,0),
\end{equation*}
the limiting support obeys
\begin{equation}
 \operatorname{supp}\mathcal P_H
 \subseteq
 [-2\sqrt{V_H},\,2\sqrt{V_H}].
 \label{eq:sm-support}
\end{equation}
Thus summable real-space correlations imply a finite
spectral edge, meaning a finite largest $|\lambda|$ beyond which the limiting density $\mathcal P_H(\lambda)$ vanishes.

At $H=1/2$, however,
\begin{equation*}
 v_H(k)\sim\frac{2}{\pi}\log\frac{1}{|k|}
 \qquad (k\to0).
\end{equation*}
Every moment of any fixed order remains finite, but high moments contain
positive contributions proportional to
\begin{equation*}
 \int_0^{k_0}
 \left[\log\frac{k_0}{k}\right]^n dk
 \propto n!\ .
\end{equation*}
The moments therefore grow too rapidly with their order for
the limiting density to have a finite edge. This explains the special
boundary behavior at $H=1/2$: every fixed moment exists, yet the support is
still unbounded.

\textit{Tail scaling.}---
For $0<H<1/2$, define
\begin{equation*}
 \alpha=1-2H>0.
\end{equation*}
Then
\begin{equation*}
 F_H(k,p)
 \sim
 A_H(k^2+p^2)^{-(1+\alpha)/2}.
\end{equation*}
Consider a large eigenvalue $|\lambda|$. The part of
Eq.~(6) of the main text responsible for such an eigenvalue comes from a small
momentum region $|k|,|p|\sim\kappa$. Over that region the covariance term is
of order $\kappa^{-\alpha}g$. Since $g\sim1/\lambda$ in the far tail, it
becomes comparable with the explicit $\lambda$ term when
\begin{equation}
 \lambda^2\sim\kappa^{-\alpha},
 \qquad
 \kappa\sim|\lambda|^{-2/\alpha}.
 \label{eq:sm-tailrescale}
\end{equation}

Only a fraction of order $\kappa$ of all momentum modes lies
in this long-wavelength region, and each contributes a resolvent of order
$1/|\lambda|$. Hence
\begin{equation}
 \mathcal P_H(\lambda)
 \sim
 \frac{\kappa}{|\lambda|}
 \propto
 |\lambda|^{-1-2/\alpha}
 =
 |\lambda|^{-1-2/(1-2H)}.
 \label{eq:sm-tail}
\end{equation}
This is the tail exponent quoted in Eq.~(12) of the main text.
The prefactor depends on the full covariance, but the exponent depends only on
the long-distance power law.

Integrating the density gives the cumulative tail
\begin{equation}
 T_H(x)\equiv\Pr(|\lambda|>x)
 \propto x^{-2/\alpha}.
 \label{eq:sm-cumtail}
\end{equation}
For finite $L$, the smallest nonzero momentum is of order $L^{-1}$ , so the tail extends only up
to the characteristic scale
\begin{equation}
 \lambda_L\sim L^{\alpha/2}
 =L^{1/2-H}.
 \label{eq:sm-tailcutoff}
\end{equation}

\textit{Moment estimators.}---
For a matrix realization $a$, define
\begin{equation*}
 t_{p,a}=\frac1L\Tr A_a^p,
 \qquad
 m_p(L)=\E\,t_{p,a}.
\end{equation*}
The second and fourth moments satisfy the exact identities
\begin{equation}
 t_2=\frac{\|A\|_F^2}{L},
 \qquad
 t_4=\frac{\|A^2\|_F^2}{L},
 \label{eq:sm-exacttraces}
\end{equation}
where $\|\cdot\|_F$ is the Frobenius norm,
\begin{equation*}
 \|B\|_F^2
 =
 \sum_{i,j}|B_{ij}|^2
 =
 \Tr(B^TB).
\end{equation*}

For the largest matrices, higher moments are obtained without diagonalizing
$A$. We use random vectors $z_b$ whose entries are independently $+1$ or
$-1$ with equal probability. Then
$ \E_z[z_i z_j]=\delta_{ij}$,
which implies
\begin{equation*}
 \E_z\!\left[z^TA^{2n}z\right]
 =
 \Tr A^{2n}.
\end{equation*}
Since
$z^TA^{2n}z=\|A^nz\|^2$,
an unbiased estimator is
\begin{equation}
 \widehat t_{2n}
 =
 \frac{1}{BL}
 \sum_{b=1}^{B}
 \|A^nz_b\|^2 .
 \label{eq:sm-probes}
\end{equation}
Thus the trace is replaced by repeated matrix--vector
multiplications, avoiding a full eigendecomposition at large $L$.

At $L=8192,16384,32768$, we use respectively $1000$, $500$, and
$300$ independent matrix realizations for each value of $H$.
The fourth-moment estimates use $B=32$ random vectors per matrix for
$H=1/8,\,1/2,\,5/8$, and $3/4$, and $B=64$ for
$H=1/4$ and $3/8$. For $H=1/4$ and $3/8$, the same $B=64$
calculations also provide the sixth and eighth moments; the $H=1/4$
results are used in Fig.~2 of the main text. Calibration against full
diagonalization is performed at $L=4096$ while increasing the number
$B$ of random probe vectors used for each matrix. The reported
uncertainty includes both matrix-to-matrix fluctuations and the
additional stochastic variation caused by using a finite number $B$
of probe vectors.

The fits in Fig.~2 of the main text use $L\ge256$ and the scaling forms
\begin{equation*}
 m_6=B_6+A_6L^{1/2},
 \qquad
 m_8=B_8+A_8L.
\end{equation*}
Unweighted least-squares fits to the central estimates give
\begin{equation*}
 \begin{aligned}
 A_6&=1.1563,\qquad &B_6&=-1.3843,\\
 A_8&=0.8342,\qquad &B_8&=54.85.
 \end{aligned}
\end{equation*}
These numerical coefficients specify the least-squares curves shown in the figure; the last displayed digits are retained for reproducibility of those curves and are not intended as a statement of statistical precision. The powers $L^{1/2}$ and $L$ in the two fit forms are fixed by Eq.~(10) of the main text; the fits determine only
the amplitudes and finite-size backgrounds. The error bars on the large-$L$ stochastic-trace data in Figs.~2 and 3 of the main text denote 95\% confidence intervals over independent matrix
realizations, including the probe-sampling uncertainty defined above.

\textit{$q$-Gaussian comparison.}---
For $H>1/4$, write
\begin{equation*}
 \chi_H=\sum_{r\ge1}R_H(r,0)^2,
\end{equation*}
so that
\begin{equation*}
 K_H=-1+4\chi_H.
\end{equation*}
A variance-normalized $q$-Gaussian with finite fourth moment has
\begin{equation*}
 K_q=\frac{6(q-1)}{7-5q}.
\end{equation*}
Matching only the kurtosis gives
\begin{equation}
 q_4(H)
 =
 \frac{28\chi_H-1}{20\chi_H+1}.
 \label{eq:sm-q4}
\end{equation}
This $q_4$ is only a convenient way of expressing the
fourth moment. It should not be confused with $q_{\rm tail}$ in
Eq.~(13) of the main text, which is obtained by matching the asymptotic tail
exponent. The two quantities probe different properties of the same spectral
density. In particular, $q_4=1$ at $H_0$, although the sixth cumulant remains
nonzero.

\textit{Density and tail observables.}---
The central-density comparison uses the full eigenspectra through $L=4096$.
To compare directly with the resolvent evaluated at
$z=\lambda+i\eta$, each numerical eigenvalue is broadened by the same
Lorentzian width:
\begin{equation}
 \mathcal P_{H,L,\eta}(\lambda)
 =
 \E\left[
 \frac{1}{\pi L}
 \sum_j
 \frac{\eta}{
 (\lambda-\lambda_j)^2+\eta^2
 }
 \right].
 \label{eq:sm-broadened}
\end{equation}
The variance-$1/2$ Gaussian and semicircle reference densities are
\begin{equation*}
 \mathcal P_{\rm G}(\lambda)
 =
 \frac{1}{\sqrt{\pi}}e^{-\lambda^2},
\end{equation*}
and
\begin{equation*}
 \mathcal P_{\rm sc}(\lambda)
 =
 \frac{1}{\pi}\sqrt{2-\lambda^2},
 \qquad
 |\lambda|\le\sqrt2.
\end{equation*}

Lorentzian broadening itself produces an artificial $|\lambda|^{-2}$ tail at
very large $|\lambda|$. For the physical tail we therefore
use the eigenvalues directly and measure the fraction lying outside
$[-x,x]$:
\begin{equation}
 T_L(x)
 =
 \E\left[
 \frac1L
 \sum_j
 \Theta(|\lambda_j|-x)
 \right],
 \label{eq:sm-tailcount}
\end{equation}
where $\Theta(y)$ is the usual Heaviside step function:
$\Theta(y)=1$ for $y>0$ and $\Theta(y)=0$ for $y<0$.
Thus $T_L(x)$ is simply the average fraction of eigenvalues
with magnitude larger than $x$. No smoothing or fitting enters its
definition.

The algebraic regime is expected above the central part of the spectrum and
below the finite-size cutoff $\lambda_L$. This cumulative
observable is therefore the natural quantity for testing the tail exponent
without contamination from the Lorentzian broadening used in
Fig.~4 of the main text.

\textit{Uncertainty and reproducibility.}---
Statistical intervals are constructed by resampling the independently generated matrices as whole statistical samples. All eigenvalues belonging to the same matrix are kept together
during resampling, preserving their correlations. Random trace probes supply
the additional stochastic error for the large-$L$ moments.

The analytical covariance sums and the self-consistent
density equation are deterministic calculations and are therefore treated
separately from ensemble uncertainty. Numerical resolution is controlled by
the momentum grid, the Lorentzian width, and the real-space cutoff used in
covariance sums. The finite periodic covariance determines the detailed
finite-size amplitudes and the central density, while the long-distance power
law determines the moment thresholds and spectral-tail exponents.

\end{document}